\documentstyle[aps]{revtex}
\begin{document}
\draft
\title{Colliding beams of light}
\author{B.V.Ivanov\thanks{%
E-mail: boyko@inrne.bas.bg}}
\address{Institute for Nuclear Research and Nuclear Energy,\\
Tzarigradsko Shausse 72, Sofia 1784, Bulgaria}
\maketitle

\begin{abstract}
The stationary gravitational field of two identical counter-moving beams of
pure radiation is found in full generality. The solution depends on an
arbitrary function and a parameter which sets the scale of the energy
density. Some of its properties are studied. Previous particular solutions
are derived as subcases.
\end{abstract}


\section{Introduction}

The gravitational field of incoherent light is induced by the
energy-momentum tensor $T_{\mu \nu }=\mu _1k_\mu k_\nu $ of pure radiation
(null dust) with energy density $\mu _1$ and travelling along the null
vector $k^\mu $. The study of steady beams and pulses of pointed radiation
has a long history. In the linear approximation to general relativity this
was done about 70 years ago \cite{one,two}. Later, Bonnor gave exact
solutions for steady and time-dependent beams \cite{three} which belong to
the well-known class of pp-waves (algebraically special solutions of Petrov
type N and no expansion) \cite{four}. He showed that beams shining in the
same direction do not interact. He also gave solutions with charged null
dust \cite{five} and spinning null dust \cite{six}. Counter-moving light
beams interact non-linearly, some explanation being proposed in Ref. \cite
{seven}. Particular exact solutions were found only recently \cite
{eight,nine}. Their Killing vectors necessarily have twist to prevent the
focusing of rays. One of these solutions has been also obtained as a
limiting case of the double-dust solution \cite{ten}. Later, the general
double-dust solution was described \cite{eleven}.

In this paper we find the stationary gravitational field of two identical
colliding beams of light in full generality. Some of its properties are
studied and the previous results are rederived as special cases.

In section 2 the properties of the energy-momentum tensor are studied, the
metric and the system of Einstein equations are written down when the energy
densities of the beams coincide. Some simplifications are done next and the
main equation is obtained. In section 3 a new radial variable is introduced
and two methods for the derivation of the general solution are given. The
explicit formulas depend on an arbitrary function and a free parameter. The
consequences of elementary flatness are studied. In section 4 an algorithm
for simple algebraic solutions is presented and the Kramer solutions are
derived as an example. The difficulties in the second method, which contains
the important case of constant energy density, are clarified. In section 5
some of the properties of the general and the particular solutions are
studied. The expansion, the acceleration and the vorticity of $k^\mu $ are
given in the most general metric. A criterion for the appearance of closed
timelike curves (CTC) is given. The static solution of Kramer does not have
CTC but its stationary counterpart does have them. Section 6 contains some
discussion. A comparison is made between the general double-dust solution
and the general solution for colliding light.

\section{Energy-momentum tensor and field equations}

We shall work in the stationary formulation of the problem when the
cylindrically symmetric metric is given by \cite{four,eight} 
\begin{equation}
ds^2=-e^{2u}\left( dt+Ad\varphi \right) ^2+e^{-2u}\left[ e^{2k}\left(
dx^2+dz^2\right) +W^2d\varphi ^2\right]  \label{one}
\end{equation}
and there is only radial $x$-dependence. The static metric may be obtained
by complex substitution; $t\rightarrow iz,z\rightarrow it,A\rightarrow iA$.
The energy-momentum tensor of the two beams reads 
\begin{equation}
T_{\mu \nu }=\mu _1k_\mu k_\nu +\mu _2l_\mu l_\nu ,  \label{two}
\end{equation}
\begin{equation}
k^\mu k_\mu =0,\qquad l^\mu l_\mu =0,  \label{three}
\end{equation}
\begin{equation}
k^\mu l_\mu =-1.  \label{four}
\end{equation}
Here $\mu _i$ are the energy density profiles of the beams and $k^\mu $, $%
l^\mu $ are their 4-velocities, which are null and normalized as in Eq. (4).
As will be shown later, there is only one solution with the minimal set of
two non-zero components. On the other hand, naturally $k^x=l^x=0$.
Therefore, let us introduce 
\begin{equation}
k^t=l^t=V,\qquad k^\varphi =l^\varphi =\Phi ,\qquad k^z=-l^z=Z  \label{five}
\end{equation}
and $T=V+A\Phi $. The beams move in the positive or negative $z$-direction.
For stationary metrics they should have the same twist $\Phi $, otherwise
the two conditions in Eq. (3) lead to contradiction. In the static picture
the twists are opposite. The obvious relation $T_\mu ^\mu =0$ and Eq. (4)
give 
\begin{equation}
e^{2u}T^2-e^{-2u}W^2\Phi ^2=e^{2\left( k-u\right) }Z^2,  \label{six}
\end{equation}
\begin{equation}
e^{2u}T^2-e^{-2u}W^2\Phi ^2+e^{2\left( k-u\right) }Z^2=1.  \label{seven}
\end{equation}
Their combinations express $Z$ and $\Phi $ in terms of the other quantities 
\begin{equation}
Z=\frac 1{\sqrt{2}}e^{u-k},  \label{eight}
\end{equation}
\begin{equation}
W^2\Phi ^2=e^{2u}\left( e^{2u}T^2-\frac 12\right) .  \label{nine}
\end{equation}

The covariant conservation of the energy-momentum tensor (which follows from
the field equations) reads 
\begin{equation}
T_{\nu ;\mu }^\mu =\frac 1{\sqrt{-g}}\frac{\partial \left( T_{\,\nu }^\mu 
\sqrt{-g}\right) }{\partial x^\mu }-\frac 12\left( g_{\mu \lambda }\right)
_\nu T^{\mu \lambda }=0,  \label{ten}
\end{equation}
where $g$ is the metric's determinant. In our case $T_\nu ^r=0$ and the
first term vanishes. The second yields 
\begin{equation}
\left( e^{2u}\right) ^{\prime }V^2+2\left( Ae^{2u}\right) ^{\prime }V\Phi
-\left( e^{-2u}W^2-e^{2u}A^2\right) ^{\prime }\Phi ^2=2\left( k-u\right)
^{\prime }e^{2\left( k-u\right) }Z^2,  \label{eleven}
\end{equation}
where $^{\prime }$ means $x$-derivative. The restriction $\Phi =0$
immediately gives the particular solution $k=2u$, as announced above.

To write the Einstein equations we use the combinations of Ricci tensor
components utilized in Refs. \cite{eleven,twelve,thirteen,fourteen}. We
shall work, however, with curved mixed components $R_\nu ^\mu $. The only
non-trivial ones are the diagonal components and $R_\varphi ^t$, $%
R_t^\varphi $. The Einstein equations for the corresponding $T_\nu ^\mu $
components give after some rearrangement 
\begin{equation}
\frac{W^{\prime \prime }}W=8\pi \rho e^{2\left( k-u\right) },  \label{twelve}
\end{equation}
\begin{equation}
2\frac{k^{\prime }W^{\prime }}W-2u^{\prime 2}+\frac{e^{4u}A^{\prime 2}}{2W^2}%
=0,  \label{thirteen}
\end{equation}
\begin{equation}
u^{\prime \prime }+\frac{u^{\prime }W^{\prime }}W+\frac{e^{4u}A^{\prime 2}}{%
2W^2}=16\pi \rho e^{2k}T^2,  \label{fourteen}
\end{equation}
\begin{equation}
k^{\prime \prime }+u^{\prime 2}+\frac{e^{4u}A^{\prime 2}}{4W^2}=16\pi \rho
e^{2k-4u}W^2\Phi ^2,  \label{fifteen}
\end{equation}
\begin{equation}
\left( \frac{e^{4u}A^{\prime }}W\right) ^{\prime }=-32\pi \rho e^{2k}W\Phi T.
\label{sixteen}
\end{equation}
Here $2\rho =\mu _1+\mu _2$. Units are used with $G=c=1$. The equations for
the rest $T_\nu ^\mu $ components are satisfied either identically or when $%
\left( \mu _1-\mu _2\right) Z=0$. Therefore, we accept in the following that 
$\mu _1=\mu _2=\rho $. There are 7 equations (8,9,12-16) for 8 unknowns $%
u,k,W,A,\rho ,V,\Phi ,Z$. Eq. (13) gives an expression for $A^{\prime }$%
\begin{equation}
A^{\prime }=\pm 2e^{-2u}W\sqrt{B},\qquad B=u^{\prime 2}-\frac{k^{\prime
}W^{\prime }}W,  \label{seventeen}
\end{equation}
which allows to eliminate $A^{\prime }$ from Eqs. (14-16). The difference
between Eqs. (14,15) leads to a relation between $k-u$ and $W$%
\begin{equation}
k^{\prime }=u^{\prime }+\frac{1-W^{\prime }}W,\qquad e^{k-u}=\frac{e^{\int 
\frac{dx}W}}W.  \label{eighteen}
\end{equation}
An unimportant constant has been set to $1$ in the first of these equations
by rescaling the $x$-coordinate. Now Eqs. (12,18) show that $\rho $ can be
expressed through $W$%
\begin{equation}
8\pi \rho =WW^{\prime \prime }e^{-2\int \frac{dx}W}  \label{nineteen}
\end{equation}
or $W$ can be found as a solution of a rather complicated equation if $\rho $
was given 
\begin{equation}
\left( WW^{\prime \prime }\right) ^{\prime }-2W^{\prime \prime }-\left( \ln
\rho \right) ^{\prime }WW^{\prime \prime }=0.  \label{twenty}
\end{equation}
Eq. (8) defines $Z$ in terms of $W$, Eqs. (14,9) express $T$ and $\Phi $ in
terms of $u$ and $W$. Thus all characteristics of the problem have
expressions as functions of $W$ or $u$ and $W$. The arbitrary function,
however, can be only one and a relation between them follows from the
crucial Eq. (16). While Eqs. (14,15) are not sensitive to the sign of $%
A^{\prime }$ in Eq.(17), Eq.(16) should be solved for both signs of $%
A^{\prime }$, leading to different solutions. They are encompassed by taking
the squares of both sides of Eq. (16) and inserting $T^2$, $\Phi ^2$ and $%
A^{\prime }$ from Eqs. (14,15,17) to obtain 
\begin{equation}
\left( 4u^{\prime }B+B^{\prime }\right) ^2=4B\left( u^{\prime \prime }+\frac{%
u^{\prime }W^{\prime }}W+2B\right) \left( u^{\prime \prime }+\frac{u^{\prime
}W^{\prime }}W+2B-\frac{W^{\prime \prime }}W\right) .  \label{twentyone}
\end{equation}
This equation involves only $u$ and $W$ due to Eqs. (17,18). From the
definition of $T$ it follows that if $A$, $\Phi $, $T$ is a solution of Eq.
(16), so is $-A$, $-\Phi $, $T$. In other words, the equations are invariant
to a change in the direction of dragging and the sign of the twist of the
pure radiation's 4-velocity.. This is different from the vacuum case \cite
{fifteen} and the case of perfect fluid \cite{thirteen}, where the r.h.s. of
Eq. (16) vanishes and the direction of rotation is not significant. In our
case the two signs of $A$ (or $\Phi $) lead to a double valuedness of the
energy density $\rho $, as will be seen from Eqs. (29, 37) below.

\section{The general solution}

Let us introduce the new radial variable $p=W^{\prime }$ and denote the $p$%
-derivatives with a subscript. Eq. (12) shows that this is possible when $%
\rho \neq 0$. Let us introduce also the important function $X$%
\begin{equation}
X=WW^{\prime \prime }=\frac{pW}{W_p}.  \label{twentytwo}
\end{equation}
Another fundamental quantity is $f$, introduced by 
\begin{equation}
u^{\prime }=\frac fW.  \label{twentythree}
\end{equation}
Then $B$ becomes 
\begin{equation}
B=\frac b{W^2},\qquad b=f^2-pf+p^2-p,  \label{twentyfour}
\end{equation}
where $b>0$ in order to have a real $A^{\prime }$. This constrains to some
extent the arbitrary function $f$. After the use of the last three equations
the main Eq. (21) turns into a quadratic equation for $X$%
\begin{equation}
C_1X^2+C_2X+C_3=0,  \label{twentyfive}
\end{equation}
\begin{equation}
C_1=b_p^2-4bf_p\left( f_p-1\right) ,  \label{twentysix}
\end{equation}
\begin{equation}
C_2=4b\left[ \left( 2f-p\right) b_p-2b\left( 2f_p-1\right) \right] ,
\label{twentyseven}
\end{equation}
\begin{equation}
C_3=4b^2\left[ \left( 2f-p\right) ^2-4b\right] .  \label{twentyeight}
\end{equation}

Every function $f$ corresponds to two solutions 
\begin{equation}
X_{a,b}=\frac{-C_2\pm \sqrt{D}}{2C_1},\qquad D=C_2^2-4C_1C_3.
\label{twentynine}
\end{equation}
The coefficients $C_i$ are simplified when the definition of $b$ is used 
\begin{equation}
C_1=p\left( 4-3p\right) f_p^2+2\left( 3pf-2f-p\right) f_p+\left(
2p-1-f\right) ^2,  \label{thirty}
\end{equation}
\begin{equation}
C_2=4b\left[ p\left( 4-3p\right) f_p+3pf-2f-p\right] ,  \label{thirtyone}
\end{equation}
\begin{equation}
C_3=4b^2p\left( 4-3p\right) .  \label{thirtytwo}
\end{equation}
It is interesting that only 4 combinations of $p$ and $f$ enter these
expressions. Another surprising fact is that $D$ factors out into terms that
do not contain $f_p$%
\begin{equation}
D=16b^2\left[ \left( 3pf-2f-p\right) ^2-p\left( 4-3p\right) \left(
2p-1-f\right) ^2\right] =64b^3\left( p-1\right) \left( 3p-1\right) .
\label{thirtythree}
\end{equation}
$X$ must be real, hence $D\geq 0$. This restricts the $p$ coordinate outside
the interval $\left( 1/3,1\right) $. The position of the axis is at $p=1$,
because $W\left( x\right) \rightarrow x$ when $x\rightarrow 0$. Thus the
realistic interval is given by $p\geq 1$. It is also obvious that Eq. (25)
is quadratic with respect to $f_p$%
\begin{eqnarray}
&&p\left( 4-3p\right) X^2f_p^2+\left[ 2\left( 3pf-2f-p\right) X^2+4p\left(
4-3p\right) bX\right] f_p+\left( 2p-1-f\right) ^2X^2  \nonumber \\
+4\left( 3pf-2f-p\right) bX+4p\left( 4-3p\right) b^2 &=&0.
\label{thirtyfour}
\end{eqnarray}

After these remarks we are ready to present the general solution of the
problem of two colliding light beams. There are two methods to do this. The
easier one is to choose an arbitrary function $f\left( p\right) $ and find
the two $X$ from the root formula (29). Then Eqs. (23,18,22,17,19,8,9 and
14) yield respectively $u,k,W,A,\rho ,Z,\Phi $ and $V$%
\begin{equation}
u=\int \frac fXdp,\qquad k=\int \frac{f+1-p}Xdp+k_0,  \label{thirtyfive}
\end{equation}
\begin{equation}
W=W_0e^{\int \frac pXdp},\qquad A=2\int_1^p\frac{\sqrt{b}}XWe^{-2u}dp,
\label{thirtysix}
\end{equation}
\begin{equation}
\rho =\rho _0Xe^{-2\int \frac{dp}X-2k_0},\qquad Z=\frac 1{\sqrt{2}}e^{\int 
\frac{p-1}Xdp-k_0},  \label{thirtyseven}
\end{equation}
\begin{equation}
T^2=\frac{2b+f_pX}{2X}e^{-2u},  \label{thirtyeight}
\end{equation}
\begin{equation}
\Phi ^2=\frac{2b+\left( f_p-1\right) X}{2W_0^2X}e^{2u-2\int \frac pX%
dp},\qquad V=T-A\Phi .  \label{thirtynine}
\end{equation}
The radial metric component $g_{xx}$ becomes $g_{pp}$%
\begin{equation}
g_{pp}=e^{2\left( k-u\right) }\frac{W^2}{X^2}.  \label{forty}
\end{equation}
One can further pass from $p$ to the physical length $L$, so that $g_{LL}=1$%
\begin{equation}
L=\int \sqrt{g_{pp}}dp=W_0e^{\int \frac{dp}X}.  \label{fortyone}
\end{equation}
The constant $W_0$ is positive by definition. The limits of the integral for 
$A$ and the integration constant $k_0$ are determined by the elementary
flatness condition 
\begin{equation}
\lim\limits_{x\rightarrow 0}\frac{e^{u-k}}x\left(
e^{-2u}W^2-e^{2u}A^2\right) ^{1/2}=1.  \label{fortytwo}
\end{equation}
$W$ and $A$ should vanish at the axis. One can see from Eq. (17) that $%
A=o\left( W\right) $ when $x\rightarrow 0$ even if there is a pole in $B$.
Therefore, it can be neglected in Eq. (42) like in many other problems \cite
{eleven,thirteen,fourteen}. The function $f$ (and $b$) also should vanish at
the axis to keep $u^{\prime }$ regular. Using the l'H\^opital rule and
passing to $p$, Eq. (42) becomes $\left( 2p-f-1\right) e^{-k}\rightarrow 1$
when $p\rightarrow 1$. This gives $k\left( 1\right) =0$ which fixes $k_0$.
Inserting $W$ and $\rho $ from Eqs. (36,37) into Eq. (12) supplies the
relation 
\begin{equation}
8\pi \rho _0W_0^2=1.  \label{fortythree}
\end{equation}
It indicates that $\rho _0$ is positive. Eq. (43) has been used in Eqs.
(38,39). Eqs. (37,43) show that $X>0$, which places another restriction on $f
$. Furthermore, the r.h.s. of Eqs. (38,39) should be positive, leading to
the inequality $2b+\left( f_p-1\right) X\geq 0$. Finally, we have kept $A$
positive in Eq. (36), allowing for a change in the sign of $\Phi $ instead.
It is determined by the sign of the l.h.s. of Eq. (16), equivalent to the
sign of $\left( 4f-2p\right) b+b_pX$.

The second, more natural but more difficult method, is to provide the sum of
the density profiles $2\rho $. Then from Eq. (20) or (37) it follows that $X$
satisfies a linear first-order differential equation 
\begin{equation}
X_p-\left( \ln \rho \right) _pX-2=0.  \label{fortyfour}
\end{equation}
Its boundary condition states that $X\left( 1\right) =0$, due to $W\left(
1\right) =0$. Its solution is 
\begin{equation}
X=2\rho \int_1^p\frac{dp}\rho .  \label{fortyfive}
\end{equation}
The next step is to determine $f$ from Eq. (34). This is, in fact, a
first-order, highly non-linear equation, whose solutions seem to exist only
in numerical form. Its boundary condition is $f\left( 1\right) =0$. After $X$
and $f$ are found, the rest of the task is fulfilled by Eqs. (35-40) where
the two methods coincide.

The general solution thus depends on one arbitrary function $f$ (satisfying
several positivity requirements) or a positive $\rho $ and the free
parameter $\rho _0$. The situation is similar to the general double-dust
solution \cite{eleven}. The constant $\rho _0$ sets the scale of the energy
density. The form of the solution is determined by Eq. (25) or its
equivalent Eq. (34). They do not contain $\rho _0$, $W_0$ or any other
parameters. An explicit solution needs the explicit calculation of only
three integrals, $\int \frac{dp}X$, $\int \frac{pdp}X$ and $\int \frac{fdp}X$%
.

\section{Particular solutions}

Let us search for some simple concrete solutions. Following the first
method, let us choose $f$ in such a way that $D$ becomes a square leading to
an algebraic $X$. According to Eq. (33) we must have 
\begin{equation}
b=\left( p-1\right) \left( 3p-1\right) b_1^2,  \label{fortysix}
\end{equation}
where $b_1$ is a polynomial. Eq. (24) shows that the simplest possibility is
constant $b_1$ and linear $f$. There are four such cases 
\begin{equation}
f_1=1-p,\qquad k=2u,  \label{fortyseven}
\end{equation}
\begin{equation}
f_2=\frac{p-1}2,\qquad k=-u,  \label{fortyeight}
\end{equation}
\begin{equation}
f_3=\frac{p+1}2,\qquad k=3u-2\ln W,  \label{fortynine}
\end{equation}
\begin{equation}
f_4=2p-1,\qquad k=\ln W.  \label{fifty}
\end{equation}
Then Eq. (29) yields 8 solutions for $X$%
\begin{equation}
X_{1a}=\left( p-1\right) \left( 3p-1\right) ,\qquad X_{1b}=\frac{p\left(
p-1\right) \left( 3p-1\right) \left( 4-3p\right) }{3p^2-4p+2},
\label{fiftyone}
\end{equation}
\begin{equation}
X_{2a}=\frac{\left( p-1\right) \left( 3p-1\right) \left( 4-3p\right) }{6p-5}%
,\qquad X_{2b}=\frac{p\left( p-1\right) \left( 3p-1\right) }{2p-1},
\label{fiftytwo}
\end{equation}
\begin{equation}
X_{3a}=-X_{2b},\qquad X_{3b}=-X_{2a},  \label{fiftythree}
\end{equation}
\begin{equation}
X_{4a}=-X_{1b},\qquad X_{4b}=-X_{1a}.  \label{fiftyfour}
\end{equation}
The first four coincide with solutions I-IV from Ref. \cite{nine}. As was
mentioned already, realistic solutions have positive $X,$ at least in some
region around the axis. Setting $p=R^2+1$ we see that $X_{1a}$ and $X_{2b}$
are positive everywhere, while $X_{1b}$ and $X_{2a}$ are positive for $%
1<p<4/3$. These solutions are suitable for interior solutions. The other
four solutions are physically unrealistic. $X_{4b}$ and $X_{3a}$ are
negative everywhere, while $X_{4a}$ and $X_{3b}$ are negative on the axis
and in its vicinity. This fact was noticed in Ref. \cite{nine}.

We have shown that when $\Phi =0$, $k=2u$. The opposite is not true since $%
X_{1b}$ leads to non-vanishing $\Phi $. Thus the solution with $f_1$ and $%
X_1 $ is the simplest possible solution. It was thoroughly investigated \cite
{eight,nine,ten}. It is worth to point out how it arises from the general
formalism. When $k=2u$, $u$ also picks up an integration constant $k_0/2$,
while the one in $\rho $ becomes $-k_0$. Eqs. (35-40) give 
\begin{equation}
e^{2u}=Y^{-2/3},\qquad e^{2k}=Y^{-4/3},\qquad e^{k_0}=2^{2/3},
\label{fiftyfive}
\end{equation}
\begin{equation}
W^2=W_0^2R^2Y^{-1/3},\qquad A=2^{1/3}W_0R^2,  \label{fiftysix}
\end{equation}
\begin{equation}
\rho =2^{4/3}\rho _0Y^2,\qquad \Phi =0,  \label{fiftyseven}
\end{equation}
\begin{equation}
Z=V=T=\frac 1{\sqrt{2}}Y^{1/3},  \label{fiftyeight}
\end{equation}
where 
\begin{equation}
Y=\frac 12\left( 3p-1\right) =1+\frac 32R^2.  \label{fiftynine}
\end{equation}
These results coincide with the results of Refs. \cite{eight,nine} when the
identifications 
\begin{equation}
8\pi \rho _0=\frac{2^{2/3}}3\lambda ^2,\qquad R=\sqrt{\frac 23}\lambda r
\label{sixty}
\end{equation}
are made. The free parameter $\lambda $ reflects the free scale of the
energy density.

Eq. (46) sets a combinatorial problem from polynomial algebra, which
probably has additional solutions with more complex $f$. Furthermore, when $%
f $ and $X$ are given analytically, the integrals in the metric may be
calculated numerically and a solution, partially explicit, partially
numerical, obtained.

The second method for generating solutions also has an analytical start by
choosing $\rho $. Then $X$ is found comparatively easy from Eq. (45). Here
are four examples 
\begin{equation}
\rho _1=const,\qquad X_1=2\left( p-1\right) ,  \label{sixtyone}
\end{equation}
\begin{equation}
\rho _2=p^{-\alpha },\qquad X_2=\frac 2{\alpha +1}\left( p-p^{-\alpha
}\right) ,  \label{sixtytwo}
\end{equation}
\begin{equation}
\rho _3=c-p,\qquad X_3=2\left( c-p\right) \ln \frac{c-1}{c-p},
\label{sixtythree}
\end{equation}
\begin{equation}
\rho _4=e^{-p},\qquad X_4=2\left( 1-e^{1-p}\right) .  \label{sixtyfour}
\end{equation}
The constants fulfil $c>1,\alpha >0$. The first case is the simplest from a
physical point of view - collision of beams of constant energy density up to
some $p_0$, where a matching to the vacuum stationary solution should be
done. Unfortunately, even for such simple $X$ Eq. (34) is so complicated
that we have been unable to solve it even numerically. This hopefully can be
done on a more powerful computer platform.

\section{Some properties of the general solution}

We have calculated with the help of GRTensor the characteristics of the
vector field $k^\mu $ in the general stationary cylindrically symmetric
metric (1). The shear's expression is too lengthy to be displayed. The
expansion $\Theta $, the acceleration $a^\mu $ and the vorticity vector $%
w^\mu $ read 
\begin{equation}
\Theta =W^{-1}\left[ \left( W\Phi \right) ^{\prime }-2W\Phi \left( u^{\prime
}-k^{\prime }\right) \right] ,  \label{sixtyfive}
\end{equation}
\begin{equation}
a^t=\Phi \left( V^{\prime }+2u^{\prime }V+\frac{e^{4u}VAA^{\prime }}{W^2}%
\right) ,  \label{sixtysix}
\end{equation}
\begin{equation}
a^\varphi =-\frac{e^{4u}V\Phi A^{\prime }}{W^2},  \label{sixtyseven}
\end{equation}
\begin{equation}
a^x=u^{\prime }e^{4u-2k}V^2+\left( u^{\prime }-k^{\prime }\right) \left(
Z^2-\Phi ^2\right) +\Phi \Phi ^{\prime },  \label{sixtyeight}
\end{equation}
\begin{equation}
a^z=\Phi \left[ 2\left( k^{\prime }-u^{\prime }\right) Z+Z^{\prime }\right] ,
\label{sixtynine}
\end{equation}
\begin{equation}
w^t=\frac{e^{2u}}{2W}\left[ 2\left( 2u^{\prime }-k^{\prime }\right)
AVZ+ZAV^{\prime }+ZA^{\prime }V-Z^{\prime }AV\right] ,  \label{seventy}
\end{equation}
\begin{equation}
w^\varphi =-\frac{e^{2u}}{2W}\left[ 2\left( 2u^{\prime }-k^{\prime }\right)
VZ+V^{\prime }Z-VZ^{\prime }\right] ,  \label{seventyone}
\end{equation}
\begin{equation}
w^x=0,  \label{seventytwo}
\end{equation}
\begin{equation}
w^z=\frac{e^{6u-2k}A^{\prime }V^2}{2W}.  \label{seventythree}
\end{equation}
When $\Phi =0$ these formulas coincide with those from Ref. \cite{eleven}.
Both $\Theta $ and $a^\mu $ are proportional to $\Phi $ and vanish in this
case except for $a^x$, which becomes 
\begin{equation}
a^x=u^{\prime }e^{4u-2k}V^2+\left( u^{\prime }-k^{\prime }\right) Z^2.
\label{seventyfour}
\end{equation}
It also vanishes for the general double-dust solution and for the only case $%
k=2u$ of colliding light with $\Phi =0$. The motion of the source is
therefore geodesic. When $\Phi \neq 0$ Eq. (8) shows that 
\begin{equation}
a^z=\left( k^{\prime }-u^{\prime }\right) \Phi Z,  \label{seventyfive}
\end{equation}
which vanishes only in the vacuum case $k=u$. The expansion does not vanish
automatically when the general solution is inserted in Eq. (65). Vorticity
does not depend on $\Phi $ at all and Eqs. (70-73) coincide with the first
relations in Eqs. (28-30) from Ref. \cite{eleven}.

It is well known that CTC exist when $g_{\varphi \varphi }<0$. We have 
\begin{equation}
g_{\varphi \varphi }=e^{2u}\left( e^{-4u}W^2-A^2\right)  \label{seventysix}
\end{equation}
and the general solution possesses CTC when 
\begin{equation}
e^{-2u}W<\int_1^p\frac{2\sqrt{b}}Xe^{-2u}Wdp.  \label{seventyseven}
\end{equation}
For the concrete solution (55-60) this yields 
\begin{equation}
r^2\left( 1+\lambda ^2r^2\right) ^{-2/3}\left( 1-\frac 13\lambda
^2r^2\right) <0,  \label{seventyeight}
\end{equation}
which is satisfied when $r>\sqrt{3}/\lambda $, so that there are CTC. In the
static case $Y=1-\lambda ^2r^2$ and inequality (78) becomes 
\begin{equation}
r^2Y^{-2/3}\left( 1+\frac 13\lambda ^2r^2\right) <0.  \label{seventynine}
\end{equation}
The l.h.s. is always positive and consequently there are no CTC. This is
also true for the exterior solution - the Levi Civita metric. In the
stationary case this role is played by the Lewis solution \cite{fifteen}. It
possesses three distinct classes and only the Weyl class is locally, but not
globally static. It seems possible to induce the Lewis class of the exterior
solution when $\rho _0$ is big enough. This class contains CTC. This
discussion of exterior solutions is valid also for the general solution.
However, the vacuum solution is not a subcase of the general solution. It
has $\rho =0$, $W=x$, so that $p$ is not a variable but a constant.

\section{Discussion}

We have obtained in this paper the general cylindrically symmetric and
stationary (no time dependence) metric when $T_{\mu \nu }$ is given by Eqs.
(2-4) and $\mu _1=\mu _2$. The components of the null vectors and the energy
density are found simultaneously with the metric. Although the solution
looks quite complicated, minimal assumptions have been made along its
derivation. Diagonal metric does not allow colliding solutions, hence, the
reason for the appearance of $A$. It prevents the focusing of rays and the
appearance of singularity \cite{eight,nine}. In the case of dust two
non-zero components of $k^\mu $ are enough for a general solution with
arbitrary energy density \cite{eleven}. In the case of light they give a
solution with specific density profile and a third component $\Phi $ is
necessary. In the dust case the field equations decouple and can be solved
one by one, giving, however, an involved expression for the energy density
in terms of the arbitrary function. In the light case there is an additional
quadratic non-linearity, because the square of Eq. (16) must be taken in
order to encompass all solutions. It reflects further in the quadratic
algebraic Eq. (25) for $X$ or the quadratic differential Eq. (34) for $f_p$.
One can choose arbitrary positive $\rho $ and find $X$ easily. The
expression for $f$ and for those parts of the metric depending on it,
however, can be only numerical. The other option is to choose $f$, but then
there are several inequalities to be satisfied for a realistic solution.

An important property is the traditional appearance of CTC in rotating
cylindrically symmetric solutions, which forms one of the differences
between the stationary and the general static solution. This takes place
both in the interior and the exterior.

In conclusion, this paper can be considered as a further step in the study
of the collision of the gravitational fields of beams of coherent or
incoherent light. The award motivating the pursuit of this issue is the
existence of singularity-free colliding solutions \cite{eight}.

\end{document}